\documentclass[twocolumn, letterpaper,10pt]{IEEEtran}

\usepackage{cite}
\usepackage{hyperref}
\usepackage{etex}
\usepackage{graphicx}
\usepackage{xcolor}
\usepackage{tikz}
\usetikzlibrary{arrows,positioning,shapes.geometric}
\usepackage{pgfplots}
\usepackage{multirow}
\usepackage{upgreek}
\usepackage{amssymb}
\usepackage{amsmath}
\usepackage{cases}
\usepackage{pifont}
\usepackage{bm}
\usepackage{amsthm}

\usepackage{tabularx}
\usepackage{booktabs}
\usepackage{filecontents}
\usepackage{subcaption}
\newcolumntype{Y}{>{\centering\arraybackslash}X}
\usepackage[noend,ruled]{algorithm2e}

\usepackage{array}
\usepackage{url}

\newcommand{\fixme}[2]{\ifx&#2&{\leavevmode\color{red}#1}\else{\leavevmode\color{red}FIXME\{}#1{\leavevmode\color{red}\}}\footnote{{\leavevmode\color{red}#2}}\PackageWarning{Fixme}{#1: #2}\fi}

\DeclareMathOperator{\PM}{PM}

\title{Design and Implementation of a Polar Codes Blind Detection Scheme}

\author{Carlo~Condo,~\IEEEmembership{Member,~IEEE}, Seyyed~Ali~Hashemi,~\IEEEmembership{Student~Member,~IEEE},
	Arash~Ardakani,
	Furkan~Ercan,~\IEEEmembership{Student~Member,~IEEE},
        Warren~J.~Gross,~\IEEEmembership{Senior~Member,~IEEE}% <-this % stops a space
\thanks{C.~Condo, S.~A.~Hashemi, A.~Ardakani, F.~Ercan and W.~J.~Gross are with the Department of Electrical and Computer Engineering, McGill University, Montr\'eal, Qu\'ebec, Canada. e-mail: carlo.condo@mail.mcgill.ca, seyyed.hashemi@mail.mcgill.ca, arash.ardakani@mail.mcgill.ca, furkan.ercan@mail.mcgill.ca, warren.gross@mcgill.ca.}% <-this % stops a space}
}

\begin{document}

\maketitle

\begin{abstract}

In blind detection, a set of candidates has to be decoded within a strict time constraint, to identify which transmissions are directed at the user equipment. Blind detection is required by the 3GPP LTE/LTE-Advanced standard, and it will be required in the $5^{\text{th}}$ generation wireless communication standard (5G) as well. Polar codes have been selected for use in 5G: thus, the issue of blind detection of polar codes must be addressed.
We propose a polar code blind detection scheme where the user ID is transmitted instead of some of the frozen bits. A first, coarse decoding phase helps selecting a subset of candidates that is decoded by a more powerful algorithm: an early stopping criterion is also introduced for the second decoding phase. Simulations results show good missed detection and false alarm rates, along with substantial latency gains thanks to early stopping. We then propose an architecture to implement the devised blind detection scheme, based on a tunable decoder that can be used for both phases. The architecture is synthesized and implementation results are reported for various system parameters. The reported area occupation and latency, obtained in 65~nm CMOS technology, are able to meet 5G requirements, and are guaranteed to meet them with even less resource usage in the latest technology nodes.

\end{abstract}

\section{Introduction} \label{sec:intro}

Blind decoding, also known as blind detection, requires the receiver of a set of bits to identify if said bits compose a codeword of a particular channel code. In 3GPP LTE/LTE-Advanced standards blind detection is used by the user equipment (UE) to receive control information related to the downlink shared channel. The UE attempts the decoding of a set of candidates, to identify if one of the candidates holds its control information. 
Blind detection will be required in the $5^{\text{th}}$ generation wireless communication standard (5G) as well: ongoing discussions are considering a substantial reduction of the time frame allocated to blind detection, from $16\mu$s to $4\mu$s. Blind detection must be performed very frequently, and given the high number of decoding attempts that must be performed in a limited time \cite{3GPP_R8}, it can lead to large implementation costs and high energy consumption. Blind detection solutions for codes adopted in previous generation standards can be found in \cite{Moosavi_GLOBECOM11,Xia_TSP14,Zhou_ENT13}.

Polar codes are a class of capacity-achieving error correcting codes, introduced by Ar{\i}kan in \cite{arikan}. They are characterized by simple encoding and decoding algorithms, and have been selected for use in 5G \cite{3gpp_polar}. In \cite{arikan}, the successive-cancellation (SC) decoding algorithm has been proposed as well. It is optimal for infinite code lengths, but its error-correction performance degrades quickly at moderate and short code lengths. In its original formulation, it also suffers from long decoding latency. SC list (SCL) decoding has been proposed in \cite{tal_list} to improve the error-correction performance of SC, at the cost of increased decoding latency. In \cite{sarkis, hashemi_SSCL, xiong_symbol, hashemi_FSSCL}, a series of techniques has been proposed, aimed at improving the decoding speed of both SC and SCL without sacrificing error-correction performance.

Blind detection of polar codes has been recently addressed in \cite{Condo_COMML17}, where a blind detection scheme fitting within 3GPP LTE-A and future 5G requirements has been proposed. It is based on a two-step scheme: a first SC decoding phase helps selecting a set of candidates, subsequently decoded with SCL. An early stopping criterion for SCL is also proposed to reduce average latency. Another recent work on polar code blind detection \cite{Giard_BD} detaches itself from 4G-5G standard requirements, and proposes a metric on which the outcome of the blind detection can be based.

In this work, we extend the blind detection scheme presented in \cite{Condo_COMML17} and its early stopping criterion by considering SCL also in the first decoding phase, and provide improved detection accuracy results. We then propose an architecture to implement the blind detection scheme: it relies on an SCL decoder with tunable list size, that can be used for both the first and second decoding stages. The architecture is synthesized and implementation results are reported for various system parameters.

The rest of the paper is organized as follows. Section~\ref{sec:prel} introduces background information on polar codes and blind detection. Section~\ref{sec:blind} details the proposed blind detection scheme, and provides simulation results to evaluate its performance. The architecture of the blind detection system is detailed in Section~\ref{sec:HW}, and implementation results are given in Section~\ref{sec:impl}. Finally, Section~\ref{sec:conc} draws the conclusion.

\section{Preliminaries} \label{sec:prel}

\subsection{Polar Codes} \label{sec:prel:PC}

A polar code $\mathcal{P}(N,K)$ is a linear block code of length $N=2^n$ and rate $K/N$, and it can be expressed as the concatenation of two polar codes of length $N/2$. This is due to the fact that the encoding process is represented by a modulo-$2$ matrix multiplication as% $\mathbf{x} = \mathbf{u} \mathbf{G}^{\otimes n}$,
\begin{equation}
\mathbf{x} = \mathbf{u} \mathbf{G}^{\otimes n}\text{,}
\end{equation}
where $\mathbf{u} = \{u_0,u_1,\ldots,u_{N-1}\}$ is the input vector, $\mathbf{x} = \{x_0,x_1,\ldots,x_{N-1}\}$ is the codeword, and the generator matrix $\mathbf{G}^{\otimes n}$ is the $n$-th Kronecker product of the polarizing matrix $\mathbf{G}=\bigl[\begin{smallmatrix} 1&0\\ 1&1 \end{smallmatrix} \bigr]$. The polarization effect brought by polar codes allows to divide the $N$-bit input vector $\mathbf{u}$ between reliable and unreliable bit-channels.
The $K$ information bits are assigned to the most reliable bit-channels of $\mathbf{u}$, while the remaining $N-K$, called frozen bits, are set to a predefined value, usually $0$.
Codeword $\mathbf{x}$ is transmitted through the channel, and the decoder receives the logarithmic likelihood ratio (LLR) vector $\mathbf{y} = \{y_0,y_1,\ldots,y_{N-1}\}$.

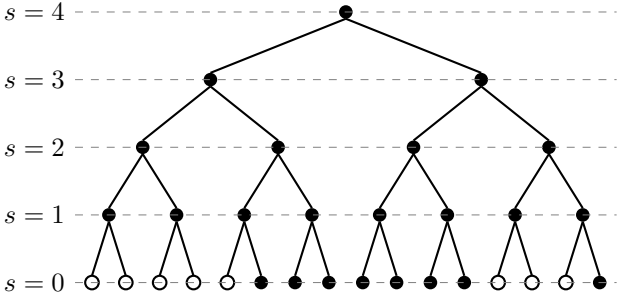
\begin{figure}
\centering
\begin{tikzpicture}[scale=1.8, thick]

\fill (0,0) circle [radius=.05];

\fill (-1,-.5) circle [radius=.05];
\fill (1,-.5) circle [radius=.05];

\fill (-1.5,-1) circle [radius=.05];
%\fill (-.5,-1) circle [radius=.06];
\fill  (-.5,-1) circle [radius=.05];
\fill  (.5,-1) circle [radius=.05];
%\fill [gray] (1.5,-1) circle [radius=.06];
\fill  (1.5,-1) circle [radius=.05];

\fill  (-1.75,-1.5) circle [radius=.05];
\fill  (-1.25,-1.5) circle [radius=.05];
\fill (-.75,-1.5) circle [radius=.05];
%\fill [white] (-.75,-1.5) circle [radius=.03];
\fill (-.25,-1.5) circle [radius=.05];
\fill  (.25,-1.5) circle [radius=.05];
\fill  (.75,-1.5) circle [radius=.05];
\fill  (1.25,-1.5) circle [radius=.05];
\fill (1.75,-1.5) circle [radius=.05];
%\fill [white] (1.75,-1.5) circle [radius=.03];

\draw (-1.875,-2) circle [radius=.05];
\draw  (-1.625,-2) circle [radius=.05];
\draw  (-1.375,-2) circle [radius=.05];
\draw  (-1.125,-2) circle [radius=.05];
\draw  (-.875,-2) circle [radius=.05];
\fill  (-.625,-2) circle [radius=.05];
\fill  (-.375,-2) circle [radius=.05];
\fill  (-.125,-2) circle [radius=.05];
\fill  (.125,-2) circle [radius=.05];
\fill  (.375,-2) circle [radius=.05];
\fill  (.625,-2) circle [radius=.05];
\fill  (.875,-2) circle [radius=.05];
\draw  (1.125,-2) circle [radius=.05];
\draw  (1.375,-2) circle [radius=.05];
\draw  (1.625,-2) circle [radius=.05];
\fill  (1.875,-2) circle [radius=.05];

\draw (0,-.05) -- (-1,-.45);
\draw (0,-.05) -- (1,-.45);

\draw (-1,-.55) -- (-1.5,-.95);
\draw (-1,-.55) -- (-.5,-.95);
\draw (1,-.55) -- (.5,-.95);
\draw (1,-.55) -- (1.5,-.95);

\draw  (-1.5,-1.05) -- (-1.75,-1.45);
\draw  (-1.5,-1.05) -- (-1.25,-1.45);
\draw (-.5,-1.05) -- (-.75,-1.45);
\draw (-.5,-1.05) -- (-.25,-1.45);
\draw  (.5,-1.05) -- (.25,-1.45);
\draw  (.5,-1.05) -- (.75,-1.45);
\draw  (1.5,-1.05) -- (1.25,-1.45);
\draw  (1.5,-1.05) -- (1.75,-1.45);

\draw  (-1.75,-1.55) -- (-1.875,-1.95);
\draw  (-1.75,-1.55) -- (-1.625,-1.95);
\draw  (-1.25,-1.55) -- (-1.375,-1.95);
\draw  (-1.25,-1.55) -- (-1.125,-1.95);
\draw  (-.75,-1.55) -- (-.875,-1.95);
\draw  (-.75,-1.55) -- (-.625,-1.95);
\draw (-.25,-1.55) -- (-.375,-1.95);
\draw (-.25,-1.55) -- (-.125,-1.95);
\draw  (.25,-1.55) -- (.125,-1.95);
\draw (.25,-1.55) -- (.375,-1.95);
\draw (.75,-1.55) -- (.625,-1.95);
\draw  (.75,-1.55) -- (.875,-1.95);
\draw  (1.25,-1.55) -- (1.125,-1.95);
\draw  (1.25,-1.55) -- (1.375,-1.95);
\draw  (1.75,-1.55) -- (1.625,-1.95);
\draw (1.75,-1.55) -- (1.875,-1.95);

\draw [very thin,gray,dashed] (-2,0) -- (2,0);
\draw [very thin,gray,dashed] (-2,-.5) -- (2,-.5);
\draw [very thin,gray,dashed] (-2,-1) -- (2,-1);
\draw [very thin,gray,dashed] (-2,-1.5) -- (2,-1.5);
\draw [very thin,gray,dashed] (-2,-2) -- (2,-2);

\node at (-2.3,0) {$s=4$};
\node at (-2.3,-.5) {$s=3$};
\node at (-2.3,-1) {$s=2$};
\node at (-2.3,-1.5) {$s=1$};
\node at (-2.3,-2) {$s=0$};

\end{tikzpicture}
\caption{Binary tree example for $\mathcal{P}(16,8)$. White circles at $s=0$ are frozen bits, black circles at $s=0$ are information bits.}
\label{fig:tree}
\end{figure}

In the seminal work on polar codes \cite{arikan}, the SC decoder is proposed. The SC-based decoding process can be represented as a binary tree search, in which the tree is explored depth first, with priority given to the left branches. Fig.~\ref{fig:tree} shows an example of SC decoding tree for $\mathcal{P}(16,8)$, where nodes at stage $s$ contain $2^s$ bits. White leaf nodes are frozen bits, while black leaf nodes are information bits.

\begin{figure}
\centering
\begin{tikzpicture}[scale=.5]

\draw [very thin,gray,dashed] (-2,0) -- (2,0);
\draw [very thin,gray,dashed] (-2,-2) -- (2,-2);
\draw [very thin,gray,dashed] (-2,-4) -- (2,-4);

\node at (-3,0) {$s+1$};
\node at (-3,-2) {$s$};
\node at (-3,-4) {$s-1$};

\fill (0,0) circle [radius=.25];
\fill  (0,-2) circle [radius=.2];
\fill  (-1.5,-4) circle [radius=.15];
\fill  (1.5,-4) circle [radius=.15];

\draw [->,very thick] (-.1,-.4) -- (-.1,-1.7) node [left,midway,rotate=0] {$\alpha$};
\draw [->,very thick] (.1,-1.7) -- (.1,-.4) node [right,midway,rotate=0] {$\beta$};

\draw [->,very thick] (-.25,-2.2) -- (-1.45,-3.75) node [left,midway,rotate=0] {$\alpha^{\text{l}}$};
\draw [->,very thick] (-1.3,-3.85) -- (-.1,-2.3) node [right,near start,rotate=0] {$\beta^{\text{l}}$};
\draw [<-,very thick] (.25,-2.2) -- (1.45,-3.75) node [right,midway,rotate=0] {$\beta^{\text{r}}$};
\draw [<-,very thick] (1.3,-3.85) -- (.1,-2.3) node [left,near start,rotate=0] {$\alpha^{\text{r}}$};

\end{tikzpicture}
\caption{Message passing in tree graph representation of SC decoding.}
\label{fig:MessagePassing}
\end{figure}
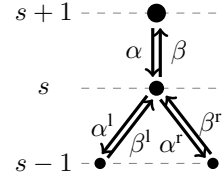

Fig.~\ref{fig:MessagePassing} portrays the message passing among SC tree nodes. Parents pass LLR values $\alpha$ to children, that send in return the hard bit estimates $\beta$. 
The left and right branch messages $\alpha^\text{l}$ and $\alpha^\text{r}$, in the hardware-friendly version of \cite{leroux}, are computed as
\begin{align}
\alpha^{\text{l}}_i = & \text{sgn}(\alpha_i)\text{sgn}(\alpha_{i+2^{s-1}})\min(|\alpha_i|,|\alpha_{i+2^{s-1}}|) \text{,} \label{eq1} \\
\alpha^{\text{r}}_i =& \alpha_{i+2^{s-1}} + (1-2\beta^\text{l}_i)\alpha_i \text{,}
\label{eq2}
\end{align}
while $\beta$ is computed as
\begin{equation}
\beta_i =
  \begin{cases}
    \beta^\text{l}_i\oplus \beta^\text{r}_i, & \text{if} \quad i < 2^{s-1} \text{,}\\
    \beta^\text{r}_{i-2^{s-1}}, & \text{otherwise},
  \end{cases}
  \label{eq3}
\end{equation}
where $\oplus$ denotes the bitwise XOR. The SC operations are scheduled according to the following order: each node receives $\alpha$ first, then sends $\alpha^\text{l}$, receives $\beta^\text{l}$, sends $\alpha^\text{r}$, receives $\beta^\text{r}$, and finally sends $\beta$.
When a leaf node is reached, $\beta_i$ is set as the estimated bit $\hat{u}_i$:
\begin{equation}
\hat{u}_i =
  \begin{cases}
    0 \text{,} & \text{if } i \in \mathcal{F} \text{ or } \alpha_{i}\geq 0\text{,}\\
    1 \text{,} & \text{otherwise,}
  \end{cases} \label{eq6}
\end{equation}
where $\mathcal{F}$ is the set of frozen bits.

The SC decoding process requires full tree exploration: however, in \cite{alamdar, sarkis} it has been shown that it is possible to prune the tree by identifying patterns in the sequence of frozen and information bits, achieving substantial speed increments. This improved SC decoding is called fast simplified SC (Fast-SSC).

%four different types of constituent codes are identified: their decoding can be carried out in a more efficient way than fully exploring the decoding tree. These codes are shown in Fig. \ref{fig:tree} as differently coloured nodes. White circles are Rate-0 nodes, whose leaves are constituted of frozen bits only, while gray circles are Rate-1 nodes, where all leaves are information bits. All leaf nodes at $s=0$ are either Rate-0 or Rate-1, being either single frozen bits or single information bits. Gray rings represent Repetition (Rep) nodes, where only the rightmost leaf is an information bit. 
%Finally, black rings represent Single Parity Check (SPC) nodes, i.e. nodes whose leaves are all information bits, except for the leftmost one. Note that SPC and Rep nodes at $s=1$ are equivalent.  %Their decoding is based on approximated formulations that involve simple computations, but cause some error-correction performance degradation. Moreover, the approximation of SPC and Rate-1 nodes is based on heuristic observations that do not necessarily remain valid if the code parameters are changed.

SC decoding suffers from modest error-correction performance with moderate and short code lengths. To improve it, the SCL algorithm was proposed in \cite{tal_list}. It is based on the same process as SC, but each time that a bit is estimated at a leaf node, both its possible values $0$ and $1$ are considered. A set of $L$ codeword candidates is stored, so that a bit estimation results in $2L$ new candidates, half of which must be discarded. To this purpose, a path metric (PM) is associated to each candidate and updated at every new estimate: the $L$ paths with the lowest PM survive. In the LLR-based SCL proposed in \cite{balatsoukas}, the hardware-friendly formulation of the PM is
\begin{align}
 \text{PM}_{{i}_l} =& \begin{cases}
    \text{PM}_{{i-1}_l}, & \text{if } \hat{u}_{i_l} = \frac{1}{2}\left(1-\text{sgn}\left(\alpha_{i_l}\right)\right)\text{,}\\
    \text{PM}_{{i-1}_l} + |\alpha_{i_l}|, & \text{otherwise,}
  \end{cases} \label{eq7}
\end{align} 
% \\
%=& \frac{1}{2}\sum_{j = 0}^{i}\text{sgn}(\alpha_{{{j}_l}})\alpha_{{{j}_l}} - (1-2\hat{u}_{j_l})\alpha_{{{j}_l}} \text{,} \label{eq7_1}   
%\end{align}
where $l$ is the path index and $\hat{u}_{i_l}$ is the estimate of bit $i$ at path $l$.
As with SC decoding, SCL tree pruning techniques relying on the identification of frozen-information bit patterns have been proposed in \cite{hashemi_SSCL,hashemi_FSSCL}, called simplified SCL (SSCL) and Fast-SSCL.

\subsection{Blind Detection}
%HERE

The physical downlink control channel (PDCCH) is used in 3GPP LTE/LTE-Advanced to transmit the downlink control information (DCI) related to the downlink shared channel. The DCI carries information regarding the channel resource allocation, transport format and hybrid automatic repeat request, and allows the UE to receive, demodulate and decode.

A cyclic redundancy check (CRC) is attached to the DCI payload before transmission. The CRC is masked according to an ID, like the radio network temporary identifier (RNTI), of the UE to which the transmission is directed, or according to one of the system-wide IDs. Finally, the DCI is encoded with a convolutional code. The UE is not aware of the format with which the DCI has been transmitted: it thus has to explore a combination of PDCCH locations, PDCCH formats, and DCI formats in the common search space (CSS) and UE-specific search space (UESSS) and attempt decoding to identify useful DCIs. This process is called blind decoding, or blind detection. For each PDCCH candidate in the search space, the UE performs channel decoding, and demasks the CRC with its ID. If no error is found in the CRC, the DCI is considered as carrying the UE control information.

%Blind detection is used by the user equipment (UE) in the physical downlink common control channel (PDCCH) in the 3GPP LTE standard to scan a set of candidate locations and decode them according to its radio network temporary identifier (RNTI), identifying if a transmission is targeting it. 
Based on LTE standard R8 \cite{3GPP_R8}, the performance specifications for the blind detection process are the following:
\begin{itemize}
 \item The DCI of PDCCH is from $8$ to $57$ bits plus $16$-bit CRC, masked by $16$-bit ID.
 \item In UESSS, a maximum of $2$ DCI formats can be sent per transmission time interval (TTI) for $2$ potential frame lengths. Therefore, $16$ candidate locations in UESSS $\rightarrow$ $32$ candidates.
 \item In CSS, a maximum of $2$ DCI formats can be sent per TTI for $2$ potential frame lengths. Therefore, $6$ candidate locations in CSS $\rightarrow$ $12$ candidates.
 \item Code length could be between $72$ and $576$ bits.
 \item Information length (including $16$-bit CRC) could be between $24$ and $73$ bits.
 \item Target signal-to-noise ratio (SNR) is dependent on the targeted block error rate (BLER): $10^{-2}$.
 \item There are two types of false-alarm scenarios: Type-1, when the UE ID is not transmitted but detected, and Type-2, when the UE ID is transmitted but another one is detected. The target false-alarm rate (FAR) is below $1.52\times 10^{-5}$.
 \item Missed detection occurs when UE ID is transmitted but not detected. The missed detection rate (MDR) is close to BLER curve.
 \item The available time frame for blind detection is $16\mu$s.
\end{itemize}

\section{Blind Detection Scheme} \label{sec:blind}

\begin{figure}[t!]
  \centering
  \begin{tikzpicture}[scale=0.9, thick]
\footnotesize
\draw [fill=white] (0,0) rectangle ++(2,3) node [pos=.5,align=center] {SCL$_1$ \\ Decoding};

\draw [->] (-1.5,2.5) -- ++(1.5,0) node [midway, above, sloped] {$0$};
\draw [->] (-1.5,2) -- ++(1.5,0) node [midway, above, sloped] {$1$};

\node at (-.75,1.5) {$\vdots$};

\draw [->] (-1.5,.5) -- ++(1.5,0) node [midway, above, sloped] {$C_1-1$};

\draw [->] (2,2.5) -- ++(.5,0);
\draw [->] (2,2) -- ++(.5,0);

\node at (2.25,1.5) {$\vdots$};

\draw [->] (2,.5) -- ++(.5,0);

\draw [fill=white] (2.5,0) rectangle ++(1.5,3) node [pos=.5,align=center] {$\PM$ \\ Sorting \\ and \\ Candidate \\ Selection};

\draw [->] (4,2) -- ++(1.5,0) node [midway, above, sloped] {$0$};

\node at (4.75,1.75) {$\vdots$};

\draw [->] (4,1) -- ++(1.5,0) node [midway, above, sloped] {$C_2-1$};

\draw [fill=white] (5.5,.5) rectangle ++(2,2) node [pos=.5,align=center] {SCL$_{\max}$ \\ Decoding};

\draw [->] (7.5,1.5) -- ++(.5,0);

\end{tikzpicture}
  \caption{Polar codes blind detection scheme.}
  \label{fig:scheme}
\end{figure}
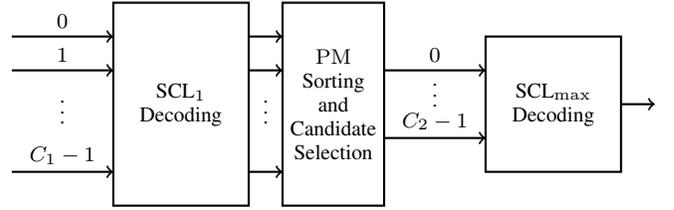

In \cite{Condo_COMML17}, polar codes have been considered within a blind detection framework, and a blind detection scheme has been proposed. Frozen bit positions are selected to instead transmit the RNTI. Fig.~\ref{fig:scheme} shows the block diagram of the devised blind detection scheme. $C_1$ candidates are received at the same time: in this case, $C_1=44$. The $C_1$ candidates are decoded with the simple SC algorithm, and a PM is obtained for each candidate, equivalent to the LLR of the last decoded bit: thanks to the serial nature of SC decoding, the LLR of the last bit can be interpreted as a reliability measure on the decoding process.
The PMs are then sorted, to help the selection of the best candidates to forward to the following decoding phase. $C_2$ candidates are in fact selected to be decoded with the more powerful SCL decoding algorithm, that guarantees a better error-correction performance, at a higher implementation complexity. The $C_2$ candidates are chosen as:
\begin{enumerate}
 \item All candidates whose ID, after the first phase, matches the one assigned to the UE. If more than $C_2$ are present, the ones with the highest PMs are selected.
 \item If free slots among the $C_2$ remain, the candidates with the smallest PMs are selected. The candidates with large PMs have higher probability to be correctly decoded: if their ID does not match the one assigned to the UE, it is probably a different one. On the other hand, candidates with small PMs have a higher chance of being incorrectly decoded, and a transmission to the UE might be hiding among them.
\end{enumerate}
After the SCL decoding phase, if one of the $C_2$ candidates matches the UE ID, it is selected, otherwise no selection is attempted.

In \cite{Condo_COMML17}, an early stopping criterion has been proposed as well, to reduce the latency and energy expenditure of the second phase of the blind detection scheme, The first phase requires the full decoding of each candidate, to identify the $C_2$ codewords that will be sent to the second phase. In the second phase, however, all codewords whose ID does not match the UE ID will be discarded. Thus, as soon as the ID is shown to be different, the decoding can be interrupted. Since SC-based decoding algorithms estimate codeword bits sequentially, the ID evaluation can be performed every time an ID bit is estimated. In case the estimated bit is different from the UE ID bit, the decoding is stopped. 

Three methods of ID bits have been described in \cite{Condo_COMML17} to choose the bits assigned to the ID:
\begin{itemize}
 \item ID mode 1: the ID bits are the $16$ most reliable bits after the $K$ information bits. 
 \item ID mode 2: the ID bits are the $16$ most reliable bits, while the $K$ information bits are the most reliable bits after the $16$ ID bits.
 \item ID mode 3: considering the order with which bits are decoded in SC-based algorithms, the ID bits are the first $16$ to be decoded among the $K+16$ most reliable bits.
\end{itemize}
The three techniques yield negligible differences in terms of error-correction performance, while ID mode 3 yields considerable advantages over mode 1 and mode 2 when early stopping is applied. In fact, since the ID bits are decoded earlier, the average percentage of estimated bits decreases, and the reduction in average latency is more substantial.

In this work, we generalize the blind detection scheme proposed in \cite{Condo_COMML17}, by considering SCL also for the first decoding phase. In particular, we consider a list sizes $L_1\ge1$ for the first decoding phase, and a list size $L_{\max}>L_1$ for the second decoding phase. It should be noted that when $L_1=1$, the blind detection scheme reverts to that of \cite{Condo_COMML17}.

\subsection{Simulation Results} \label{sec:simres}

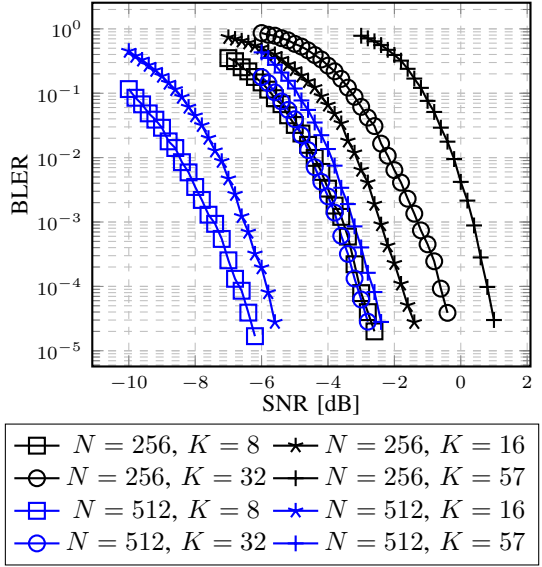
\begin{figure}[t!]
  \centering
  \begin{tikzpicture}
  \pgfplotsset{
    label style = {font=\fontsize{9pt}{7.2}\selectfont},
    tick label style = {font=\fontsize{7pt}{7.2}\selectfont},
  }

\begin{axis}[
	scale = 0.85,
    ymode=log,
    xlabel={SNR [dB]}, xlabel style={yshift=0.8em},
    ylabel={BLER}, ylabel style={yshift=-0.75em},%
    grid=both,
    ymajorgrids=true,
    xmajorgrids=true,
    %width=\columnwidth%, height=7.0cm,
    grid style=dashed,
    thick,
    mark size=3,
    legend columns=2,
    %legend pos=outer north east,
        legend to name=blerlegend
]

\addplot[
    color=black,
    mark=square,
    thick,
    mark size=3,
]
table {
-7.0 0.3513
-6.8 0.3151
-6.6 0.2531
-6.4 0.2184
-6.2 0.1789
-6.0 0.1472
-5.8 0.1137
-5.6 0.0839
-5.4 0.0676
-5.2 0.0454
-5.0 0.0322
-4.8 0.0237
-4.6 0.0158
-4.4 0.0099
-4.2 0.006
-4.0 0.0033
-3.8 0.0018
-3.6 0.00120213
-3.4 0.000593398
-3.2 0.000220759
-3.0 7.8661e-05
-2.8 3.98465e-05
-2.6 2.00092e-05
};
\addlegendentry{$N=256$, $K=8$}

\addplot[
    color=black,
    mark=star,
    thick,
    mark size=3,
]
table {
-7.0 0.7394
-6.8 0.6963
-6.6 0.653
-6.4 0.6084
-6.2 0.5444
-6.0 0.5083
-5.8 0.4457
-5.6 0.4003
-5.4 0.338
-5.2 0.2784
-5.0 0.2339
-4.8 0.1867
-4.6 0.1527
-4.4 0.1231
-4.2 0.0892
-4.0 0.0658
-3.8 0.0478
-3.6 0.034
-3.4 0.0182
-3.2 0.0119
-3.0 0.0064
-2.8 0.0041
-2.6 0.0019
-2.4 0.000912008
-2.2 0.000428043
-2.0 0.000228655
-1.8 0.00010922
-1.6 0.0000508073
-1.4 2.76603e-05 
};
\addlegendentry{$N=256$, $K=16$}

\addplot[
    color=black,
    mark=o,
    thick,
    mark size=3,
]
table {
-6.0 0.8613
-5.8 0.8105
-5.6 0.7834 
-5.4 0.7167
-5.2 0.6622
-5.0 0.5849
-4.8 0.5312
-4.6 0.4636
-4.4 0.3926
-4.2 0.3345
-4.0 0.2683
-3.8 0.2189
-3.6 0.1671
-3.4 0.1222
-3.2 0.0879
-3.0 0.0614
-2.8 0.0419
-2.6 0.0309
-2.4 0.0165
-2.2 0.0107
-2.0 0.0064
-1.8 0.0041
-1.6 0.0023
-1.4 0.00134816
-1.2 0.000741174
-1.0 0.000450545
-0.8 0.000245469
-0.6 9.23191e-05
-0.4 3.8818e-05
 };
\addlegendentry{$N=256$, $K=32$}

\addplot[
    color=black,
    mark=+,
    thick,
    mark size=3,
]
table {
-3.0 0.7814
-2.8 0.7125
-2.6 0.6401
-2.4 0.5597
-2.2 0.4726
-2.0 0.3932
-1.8 0.3209
-1.6 0.2396
-1.4 0.169
-1.2 0.1212
-1.0 0.0761
-0.8 0.0512
-0.6 0.0292
-0.4 0.0177
-0.2 0.0095
0.0 0.0042
0.2 0.00215
0.4 0.00088432
0.6 0.000281583
0.8 9.80426e-05
1.0 3.00477e-05
};
\addlegendentry{$N=256$, $K=57$}
\addplot[
    color=blue,
    mark=square,
    thick,
    mark size=3,
]
table {
-10.0 0.116
-9.8 0.0851
-9.6 0.0672	
-9.4 0.0487
-9.2 0.0389
-9.0 0.0291
-8.8 0.0179
-8.6 0.0141
-8.4 0.0085
-8.2 0.0059
-8.0 0.0035
-7.8 0.0022
-7.6 0.00129838
-7.4 0.000937207
-7.2 0.00054574
-7.0 0.000254206
-6.8 0.000131204
-6.6 8.55781e-05
-6.4 3.91505e-05
-6.2 1.68622e-05
};
\addlegendentry{$N=512$, $K=8$}
\addplot[
    color=blue,
    mark=star,
    thick,
    mark size=3,
]
table {
-10.0 0.4508
-9.8 0.3784
-9.6 0.3294
-9.4 0.2728
-9.2 0.2281
-9.0 0.1844
-8.8 0.1439
-8.6 0.1145
-8.4 0.0871
-8.2 0.061
-8.0 0.0447
-7.8 0.0317
-7.6 0.0201
-7.4 0.0128
-7.2 0.0088
-7.0 0.0047
-6.8 0.0027
-6.6 0.001223865
-6.4 0.000698429
-6.2 0.000320168
-6.0 0.000195414
-5.8 8.02856e-05 
-5.6 2.78062e-05
};
\addlegendentry{$N=512$, $K=16$}
\addplot[
    color=blue,
    mark=o,
    thick,
    mark size=3,
]
table {
-6.0 0.1783
-5.8 0.1438
-5.6 0.1032
-5.4 0.0746
-5.2 0.0552
-5.0 0.0395
-4.8 0.0223
-4.6 0.0131
-4.4 0.0071
-4.2 0.0042
-4.0 0.0025
-3.8 0.0014
-3.6 0.000609785
-3.4 0.000319303
-3.2 0.000132523
-3.0 6.25344e-05
-2.8 2.83066e-05
};
\addlegendentry{$N=512$, $K=32$}
\addplot[
    color=blue,
    mark=+,
    thick,
    mark size=3,
]
table {
-6.0 0.4311
-5.8 0.3527
-5.6 0.2769
-5.4 0.2094
-5.2 0.1563
-5.0 0.1184
-4.8 0.0773
-4.6 0.0551
-4.4 0.0351
-4.2 0.0223
-4.0 0.0136
-3.8 0.0075
-3.6 0.0034
-3.4 0.0019
-3.2 0.000855774
-3.0 0.000403076
-2.8 0.00016248
-2.6 8.17165e-05
-2.4 2.80288e-05
};
\addlegendentry{$N=512$, $K=57$}

\end{axis}
\end{tikzpicture}
    \\
  \ref{blerlegend}
  \caption{BLER curves with SCL when $L=8$.}
  \label{fig:BLER}
\end{figure}

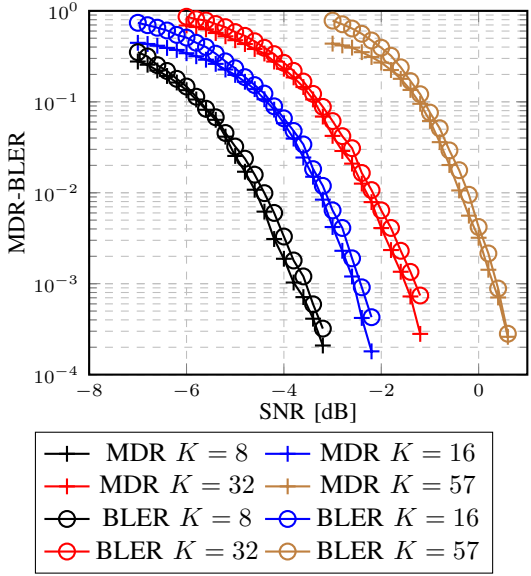
\begin{figure}[t!]
  \centering
    \begin{tikzpicture}
  \pgfplotsset{
    label style = {font=\fontsize{9pt}{7.2}\selectfont},
    tick label style = {font=\fontsize{7pt}{7.2}\selectfont},
  }

\begin{axis}[
    scale = 0.85,
    ymode=log,
    xlabel={SNR [dB]}, xlabel style={yshift=0.8em},
    ylabel={MDR-BLER}, ylabel style={yshift=-0.75em},%
    grid=both,
    ymajorgrids=true,
    xmajorgrids=true,
    %width=\columnwidth%, height=7.0cm,
    grid style=dashed,
    thick,
    mark size=3,
    legend columns=2,
     ymin=0.0001, ymax=1,
     xmin=-8, xmax=1,
    %legend pos=outer north east,
     legend to name=MDRlegend
]

\addplot[
    color=black,
    mark=+,
    thick,
    mark size=3,
]
table {
-7.0 0.2775
-6.8 0.2558
-6.6 0.2210
-6.4 0.1924
-6.2 0.1629
-6.0 0.134248
-5.8 0.10601
-5.6 0.0818
-5.4 0.0634
-5.2 0.0414
-5.0 0.0254
-4.8 0.01703
-4.6 0.0108
-4.4 0.0062
-4.2 0.0031
-4.0 0.00188
-3.8 0.00103
-3.6 0.000712
-3.4 0.000412
-3.2 0.000209
}; \label{k8}
\addlegendentry{MDR $K=8$}

\addplot[
    color=blue,
    mark=+,
    thick,
    mark size=3,
]
table {
-7.0 0.4452
-6.8 0.4324
-6.6 0.4136
-6.4 0.3912
-6.2 0.3718
-6.0 0.3422
-5.8 0.3170
-5.6 0.2904
-5.4 0.2614
-5.2 0.2264
-5.0 0.1976
-4.8 0.1661
-4.6 0.137851
-4.4 0.106731
-4.2 0.0820
-4.0 0.0582
-3.8 0.0394
-3.6 0.0244
-3.4 0.015
-3.2 0.0084
-3.0 0.0042
-2.8 0.0023
-2.6 0.0012
-2.4 0.00042
-2.2 0.0001803

};\label{k16}
\addlegendentry{MDR $K=16$}

\addplot[
    color=red,
    mark=+,
    thick,
    mark size=3,
]
table {
-6.0 0.69
-5.8 0.6644 
-5.6 0.6268
-5.4 0.571
-5.2 0.5256
-5.0 0.4756
-4.8 0.4318
-4.6 0.381
-4.4 0.3396
-4.2 0.2772
-4.0 0.231
-3.8 0.1856
-3.6 0.1454
-3.4 0.1062
-3.2 0.0688
-3.0 0.0422
-2.8 0.0288
-2.6 0.0209
-2.4 0.0126
-2.2 0.00789
-2.0 0.004089
-1.8 0.002344
-1.6 0.001353
-1.4 0.0007236
-1.2 0.0002805
};\label{k32}
\addlegendentry{MDR $K=32$}

\addplot[
    color=brown,
    mark=+,
    thick,
    mark size=3,
]
table {
-3.0 0.4364
-2.8 0.4168
-2.6 0.3944
-2.4 0.3516
-2.2 0.3110
-2.0 0.2702
-1.8 0.223604
-1.6 0.179036
-1.4 0.135059
-1.2 0.095096
-1.0 0.0615951
-0.8 0.0360179
-0.6 0.0201825
-0.4 0.0106
-0.2 0.0056
0.0 0.0032
0.2 0.001423
0.4 0.0007056
0.6 0.0002606

}; \label{k57}
\addlegendentry{MDR $K=57$}
% 
% 
% \end{axis}
% 
% \begin{axis}[
%     scale = 0.85,
%     ymode=log,
%     axis y line*=right,
%     axis x line=none,
%    xlabel={SNR [dB]}, xlabel style={yshift=0.8em},
%     ylabel={BLER},% ylabel style={yshift=-5.75em}
%     grid=both,
%     ymajorgrids=true,
%     xmajorgrids=true,
%     width=\columnwidth%, height=7.0cm,
%     grid style=dashed,
%     thick,
%     mark size=3,
%     legend style={font=\Tiny},
%     legend columns=2,
%     ymin=0.0001, ymax=1,
%     xmin=-8, xmax=1,
%     legend pos=outer north east,
%     legend to name=MDRlegend
% ]
% \addlegendimage{/pgfplots/refstyle=k8}\addlegendentry{MDR $K=8$}
% \addlegendimage{/pgfplots/refstyle=k16}\addlegendentry{MDR $K=16$}
% \addlegendimage{/pgfplots/refstyle=k32}\addlegendentry{MDR $K=32$}
% \addlegendimage{/pgfplots/refstyle=k57}\addlegendentry{MDR $K=57$}
\addplot[
    color=black,
    mark=o,
    thick,
    mark size=3,
]
table {
-7.0 0.3513
-6.8 0.3151
-6.6 0.2531
-6.4 0.2184
-6.2 0.1789
-6.0 0.1472
-5.8 0.1137
-5.6 0.0839
-5.4 0.0676
-5.2 0.0454
-5.0 0.0322
-4.8 0.0237
-4.6 0.0158
-4.4 0.0099
-4.2 0.006
-4.0 0.0033
-3.8 0.0018
-3.6 0.00120213
-3.4 0.000593398
-3.2 0.000320759
};
\addlegendentry{BLER $K=8$}
\addplot[
    color=blue,
    mark=o,
    thick,
    mark size=3,
]
table {
-7.0 0.7394
-6.8 0.6963
-6.6 0.653
-6.4 0.6084
-6.2 0.5444
-6.0 0.5083
-5.8 0.4457
-5.6 0.4003
-5.4 0.338
-5.2 0.2784
-5.0 0.2339
-4.8 0.1867
-4.6 0.1527
-4.4 0.1231
-4.2 0.0892
-4.0 0.0658
-3.8 0.0478
-3.6 0.034
-3.4 0.0182
-3.2 0.0119
-3.0 0.0064
-2.8 0.0041
-2.6 0.0019
-2.4 0.000912008
-2.2 0.000428043
};
\addlegendentry{BLER $K=16$}

\addplot[
    color=red,
    mark=o,
    thick,
    mark size=3,
]
table {
-6.0 0.8613
-5.8 0.8105
-5.6 0.7834 
-5.4 0.7167
-5.2 0.6622
-5.0 0.5849
-4.8 0.5312
-4.6 0.4636
-4.4 0.3926
-4.2 0.3345
-4.0 0.2683
-3.8 0.2189
-3.6 0.1671
-3.4 0.1222
-3.2 0.0879
-3.0 0.0614
-2.8 0.0419
-2.6 0.0309
-2.4 0.0165
-2.2 0.0107
-2.0 0.0064
-1.8 0.0041
-1.6 0.0023
-1.4 0.00134816
-1.2 0.000741174
% -1.0 0.000450545
% -0.8 0.000245469
% -0.6 9.23191e-05
% -0.4 3.8818e-05
};
\addlegendentry{BLER $K=32$}

\addplot[
    color=brown,
    mark=o,
    thick,
    mark size=3,
]
table {
-3.0 0.7814
-2.8 0.7125
-2.6 0.6401
-2.4 0.5597
-2.2 0.4726
-2.0 0.3932
-1.8 0.3209
-1.6 0.2396
-1.4 0.169
-1.2 0.1212
-1.0 0.0761
-0.8 0.0512
-0.6 0.0292
-0.4 0.0177
-0.2 0.0095
0.0 0.0042
0.2 0.00215
0.4 0.00088432
0.6 0.000281583
%0.8 9.80426e-05
%1.0 3.00477e-05
};
\addlegendentry{BLER $K=57$}

\end{axis}

\end{tikzpicture}
      \\
  \ref{MDRlegend}
  \caption{Missed detection rates after the second decoding phase with $L_1=2$, $L_{\max}=8$, and $C_2=5$. Transmissions include $C_1/2$ cases of $N_1=256$ and $C_1/2$ cases of $N_2=512$.}
  \label{fig:MD_SCL}
\end{figure}

To evaluate the effectiveness of the proposed blind detection scheme, simulations were performed. The BLER, MDR, and FAR have been measured on the additive white Gaussian noise (AWGN) channel, with binary phase-shift keying (BPSK) modulation, at the variation of different code parameters. We focused on polar codes with block lengths $N=\{256, 512\}$, since in \cite{Condo_COMML17} it has been shown that they constitute the most critical cases in terms of speed. Four information lengths $K=\{8, 16, 32, 57\}$ have been considered, while the number of ID bits has been set to $16$. The 3GPP standardization committee has decided that information bits in polar codes must be assigned to the $K$ most reliable bit-channels \cite{3gpp_polar_AH}: thus, the ID bits have been assigned according to ID mode 1. The ID values assigned to the $C_1$ candidates are randomly selected over $16$ bits. While different numbers of candidates passed to the second phase have been considered in \cite{Condo_COMML17}, we have focused here on $C_2=5$, for which a good tradeoff between accuracy and latency is found. At the same time, we set $L_{\max}=8$ and $L_1=2$: it is a representative case for which $L_{\max}$ guarantees good error-correction performance, and at which SCL decoders can be implemented with reasonable complexity.

Fig.~\ref{fig:BLER} plots the BLER curves for all the considered code lengths and rates. As expected, their error-correction performance improves as the code length increases and the code rate decreases. In Fig.~\ref{fig:MD_SCL}, the first of the metrics specific to the blind detection problem, the MDR, is depicted. The MDR can be defined as the number of missed detections divided by the number of transmissions in which the UE ID was sent. The curves in Fig.~\ref{fig:MD_SCL} have been obtained considering $C_1/2$ candidates of length $N_1=256$, and $C_1/2$ candidates of length $N_2=512$ in each transmission, with $K_1=K_2$ information bits. Together with the MDR, in Fig.~\ref{fig:MD_SCL} the BLER curves relative to the aggregate transmissions are portrayed. It can be seen that the MDR curve is always lower than the relative BLER curve.

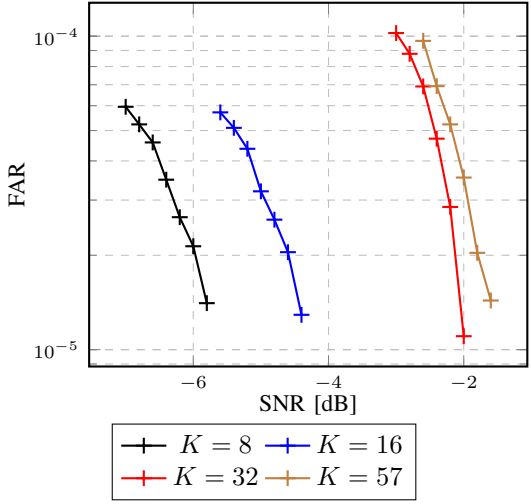
\begin{figure}[t!]
  \centering
  \begin{tikzpicture}
  \pgfplotsset{
    label style = {font=\fontsize{9pt}{7.2}\selectfont},
    tick label style = {font=\fontsize{7pt}{7.2}\selectfont},
  }

\begin{axis}[
	scale = 0.85,
    ymode=log,
    xlabel={SNR [dB]}, xlabel style={yshift=0.8em},
    ylabel={FAR}, ylabel style={yshift=-0.75em},%
    grid=both,
    ymajorgrids=true,
    xmajorgrids=true,
    %width=\columnwidth%, height=7.0cm,
    grid style=dashed,
    thick,
    mark size=3,
    legend columns=2,
    %legend pos=outer north east,
    legend to name=FARlegend
]

\addplot[
    color=black,
    mark=+,
    %dash pattern=on 1pt off 3pt on 3pt off 3pt,
    thick,
    mark size=3,
]
table {
-7.0 0.000059523
-6.8 0.0000523087
-6.6 0.0000458113
-6.4 0.0000348434
-6.2 0.000026455
-6.0 0.00002137
-5.8 0.00001407105
};
\addlegendentry{$K=8$}

\addplot[
    color=blue,
    mark=+,
    thick,
    %dash pattern=on 1pt off 3pt on 3pt off 3pt,
    mark size=3,
]
table {
-5.6 0.000057132
-5.4 0.0000509876
-5.2 0.000043722
-5.0 0.000031994
-4.8 0.00002598
-4.6 0.00002046
-4.4 0.00001291681
};
\addlegendentry{$K=16$}

\addplot[
    color=red,
    mark=+,
    thick,
    %dash pattern=on 1pt off 3pt on 3pt off 3pt,
    mark size=3,
]
table {
-3.0 0.0001023
-2.8 0.00008783
-2.6 0.000069121
-2.4 0.000047098
-2.2 0.00002852
-2.0 0.00001104
};
\addlegendentry{$K=32$}

\addplot[
    color=brown,
    mark=+,
    thick,
    %dash pattern=on 1pt off 3pt on 3pt off 3pt,
    mark size=3,
]
table {
-2.6 0.00009659
-2.4 0.00006925
-2.2 0.00005231
-2.0 0.00003540
-1.8 0.00002036
-1.6 0.00001435
};
\addlegendentry{$K=57$}

\end{axis}
\end{tikzpicture}
  \\
  \ref{FARlegend}
  \caption{False alarm rates after the second decoding phase with $L_1=2$, $L_{\max}=8$, and $C_2=5$. Transmissions include $C_1/2$ cases of $N_1=256$ and $C_1/2$ cases of $N_2=512$.}
  \label{fig:FAR}
\end{figure}

The FAR curves for the considered case study are portrayed in Fig.~\ref{fig:FAR}. The system target FAR is equivalent to the FAR obtained with a $16$-bit CRC: in 5G, a CRC of at least $16$-bits long is foreseen. Here, we evaluate the additional contribution that the proposed blind detection scheme can bring in lowering the FAR on top of the CRC. It can be seen that the FAR is kept below the $10^{-4}$ threshold at SNR values for which the BLER is still very high, and decreases as the channel conditions improve. In the blind detection method presented in \cite{Giard_BD}, the FAR increases as the MDR decreases. On the other hand, the proposed scheme allows to decrease both at the same time, thus avoiding performance limitations that could make it unappealing for 5G standard applications.
%Unlike the blind detection method presented in \cite{Giard_BD}, where the FAR increases as the MDR decreases, the proposed scheme allows to decrease both at the same time. 

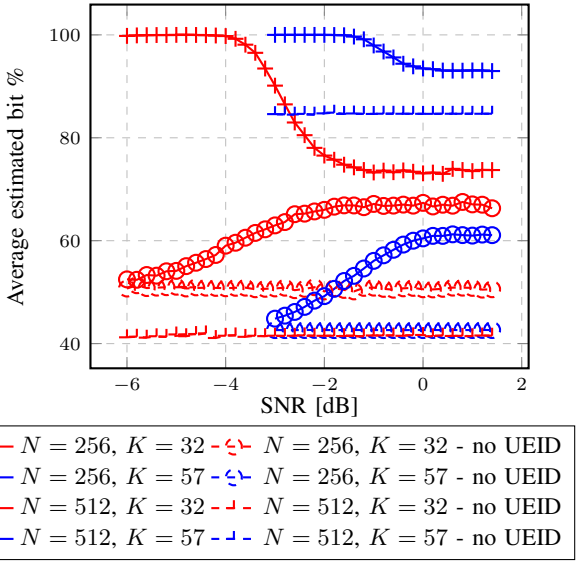
\begin{figure}[t!]
  \centering
  \begin{tikzpicture}
  \pgfplotsset{
    label style = {font=\fontsize{9pt}{7.2}\selectfont},
    tick label style = {font=\fontsize{7pt}{7.2}\selectfont},
  }

\begin{axis}[
	scale = 0.85,
    xlabel={SNR [dB]}, xlabel style={yshift=0.8em},
    ylabel={Average estimated bit \%}, ylabel style={yshift=-0.75em},%
    grid=both,
    ymajorgrids=true,
    xmajorgrids=true,
    %width=\columnwidth%, height=7.0cm,
    grid style=dashed,
    thick,
    mark size=3,
    legend columns=2,
    legend pos=outer north east,
    legend to name=ESlegend
]
% 
% \addplot[
%     color=black,
%     mark=o,
%     thick,
%     mark size=3,
% ]
% table {
% -2.4 60.4
% -2.2 60.7
% -2.0 60.7
% -1.8 60.9
% };
% \addlegendentry{$N128$ $K8$ }
% 
% \addplot[
%     color=black,
%     mark=o,
%     thick,
%     dashed,
%     mark size=3,
% ]
% table {
% -2.4 50.8
% -2.2 50.8
% -2.0 50.8
% -1.8 50.8
% };
% \addlegendentry{$N128$ $K8$ - no UEID}
% 
% 
% \addplot[
%     color=blue,
%     mark=o,
%     thick,
%     mark size=3,
% ]
% table {
% -2.4 41.9
% -2.2 42.2
% -2.0 42.3
% -1.8 42.3
% -1.6 42.7
% -1.4 43.2
% };
% \addlegendentry{$N128$ $K16$ }
% 
% \addplot[
%     color=blue,
%     mark=o,
%     thick,
%     dashed,
%     mark size=3,
% ]
% table {
% -2.4 32.6
% -2.2 32.6
% -2.0 32.6
% -1.8 32.6
% -1.6 32.6
% -1.4 32.6
% };
% \addlegendentry{$N128$ $K16$ - no UEID}

\addplot[
    color=red,
    mark=o,
    thick,
    mark size=3,
]
table {
-6.0 52.4821
-5.8 52.4021
-5.6 53.3316
-5.4 53.2027
-5.2 54.0547
-5.0 54.1337
-4.8 55.0528
-4.6 55.7135
-4.4 56.4321
-4.2 57.2277
-4.0 59.0314
-3.8 59.6019
-3.6 60.5782
-3.4 61.5095
-3.2 62.2115
-3.0 62.9746
-2.8 63.6205
-2.6 65.1471
-2.4 65.2563
-2.2 65.6846
-2.0 65.9959
-1.8 66.6119
-1.6 66.8666
-1.4 66.8479
-1.2 66.5114
-1.0 67.1444
-0.8 66.7569
-0.6 66.8181
-0.4 67.0109
-0.2 66.8138
-0.0 67.3514
0.2 66.6534
0.4 66.9891
0.6 66.8035
0.8 67.5306
1.0 67.0569
1.2 66.9946
1.4 66.303
};
\addlegendentry{\small $N=256$, $K=32$ }

\addplot[
    color=red,
    mark=o,
    thick,
    dashed,
    mark size=3,
]
table {
-6.0 50.7001
-5.8 51.3413
-5.6 50.7928
-5.4 50.8776
-5.2 50.9593
-4.6 50.9296
-5.0 51.0528
-4.8 50.7135
-4.4 50.8746
-4.2 50.8273
-4.0 50.7555
-3.8 50.8469
-3.6 50.8732
-3.4 50.5605
-3.2 50.6492
-3.0 50.7222
-2.8 50.8653
-2.6 50.3835
-2.4 50.6413
-2.2 50.9586
-2.0 50.141
-1.8 50.7785
-1.6 50.431
-1.4 50.145
-1.2 50.8307
-1.0 50.8723
-0.8 50.6376
-0.6 50.4801
-0.4 50.7441
-0.2 50.8298
-0.0 50.5578
0.2 50.532
0.4 50.647
0.6 50.8788
0.8 50.8761
1.0 50.8282
1.2 50.4919
1.4 50.5618
};
\addlegendentry{\small $N=256$, $K=32$ - no UEID}

\addplot[
    color=blue,
    mark=o,
    thick,
    mark size=3,
]
table {
-3.0 44.8442
-2.8 45.4292
-2.6 46.1549
-2.4 47.1346
-2.2 48.3509
-2.0 49.2247
-1.8 50.641
-1.6 52.1691
-1.4 53.1685
-1.2 54.5942
-1.0 56.0787
-0.8 57.1569
-0.6 58.2293
-0.4 59.2434
-0.2 59.9965
0.0 60.448
0.2 60.9378
0.4 60.9333
0.6 61.2368
0.8 61.0883
1.0 60.9909
1.2 61.1888
1.4 61.0788
};
\addlegendentry{\small $N=256$, $K=57$ }

\addplot[
    color=blue,
    mark=o,
    thick,
    dashed,
    mark size=3,
]
table {
-3.0 42.6271
-2.8 42.6284
-2.6 42.6225
-2.4 42.6432
-2.2 42.6617
-2.0 42.6374
-1.8 42.6256
-1.6 42.6385
-1.4 42.6428
-1.2 42.6444
-1.0 42.6428
-0.8 42.6244
-0.6 42.6398
-0.4 42.6118
-0.2 42.6477
0.0 42.6276
0.2 42.6468
0.4 42.6351
0.6 42.647
0.8 42.633
1.0 42.6387
1.2 42.6563
1.4 42.6348
};
\addlegendentry{\small $N=256$, $K=57$ - no UEID}

% 
% 
% \addplot[
%     color=black,
%     mark=+,
%     thick,
%     mark size=3,
% ]
% table {
% -2.4 67.7
% -2.2 67.9
% -2.0 68.2
% -1.8 68.6
% };
% \addlegendentry{$N256$ $K8$ }
% 
% \addplot[
%     color=black,
%     mark=+,
%     thick,
%     dashed,
%     mark size=3,
% ]
% table {
% -2.4 51.6
% -2.2 51.6
% -2.0 51.6
% -1.8 51.6
% };
% \addlegendentry{$N256$ $K8$ - no UEID}
% 
% 
% \addplot[
%     color=blue,
%     mark=+,
%     thick,
%     mark size=3,
% ]
% table {
% -2.4 58.8
% -2.2 61.2
% -2.0 64.4
% -1.8 66.6
% -1.6 67.9
% -1.4 68.3
% };
% \addlegendentry{$N256$ $K16$ }
% 
% \addplot[
%     color=blue,
%     mark=+,
%     thick,
%     dashed,
%     mark size=3,
% ]
% table {
% -2.4 41.4
% -2.2 41.4
% -2.0 41.4
% -1.8 41.4
% -1.6 41.4
% -1.4 41.4
% };
% \addlegendentry{$N256$ $K16$ - no UEID}

\addplot[
    color=red,
    mark=+,
    thick,
    mark size=3,
]
table {
-6.0 99.798
-5.8 99.8989
-5.6 99.9427
-5.4 99.9696
-5.2 99.9697
-4.6 99.9846
-5.0 100
-4.8 100
-4.4 99.947
-4.2 99.878
-4.0 99.7726
-3.8 99.2377
-3.6 98.0955
-3.4 96.5124
-3.2 93.463
-3.0 90.1351
-2.8 86.4955
-2.6 82.9733
-2.4 80.5055
-2.2 78.0395
-2.0 76.5435
-1.8 75.7342
-1.6 74.7719
-1.4 74.2542
-1.2 73.7394
-1.0 73.2592
-0.8 73.6259
-0.6 73.3096
-0.4 73.6611
-0.2 73.4022
-0.0 73.0856
0.2 73.2408
0.4 72.9868
0.6 73.959 
0.8 73.7416
1.0 73.5274
1.2 73.7515
1.4 73.731
};
\addlegendentry{\small $N=512$, $K=32$ }

\addplot[
    color=red,
    mark=+,
    thick,
    dashed,
    mark size=3,
]
table {
-6.0 41.221
-5.8 41.3421
-5.6 41.16
-5.4 41.4291
-5.2 41.3211
-4.6 41.8653
-5.0 41.7798
-4.8 41.4387
-4.4 41.9867
-4.2 41.0904
-4.0 41.5234
-3.8 41.6024
-3.6 41.3173
-3.4 41.5245
-3.2 41.4497
-3.0 41.5317
-2.8 41.4922
-2.6 41.5249
-2.4 41.6002 
-2.2 41.52
-2.0 41.5185
-1.8 41.5265
-1.6 41.4689
-1.4 41.5465
-1.4 41.5001
-1.2 41.4139
-1.0 41.5742
-0.8 41.5852
-0.6 41.5801
-0.4 41.5193
-0.2 41.6217
0.0 41.6763
0.2 41.4326
0.4 41.5547
0.6 41.4851
0.8 41.5115
1.0 41.4476
1.2 41.5729
1.4 41.5436
};
\addlegendentry{\small $N=512$, $K=32$ - no UEID}

\addplot[
    color=blue,
    mark=+,
    thick,
    mark size=3,
]
table {
-3.0 100
-2.8 100
-2.6 100
-2.4 100
-2.2 100
-2.0 100
-1.8 99.9752
-1.6 99.9312
-1.4 99.6718
-1.2 99.1387
-1.0 97.9433
-0.8 96.7679
-0.6 95.6187
-0.4 94.3968
-0.2 93.8846
0.0 93.4722
0.2 93.3382
0.4 93.073
0.6 93.045
0.8 93.0482
1.0 93.0918
1.2 93.0829
1.4 92.9719
};
\addlegendentry{\small $N=512$, $K=57$ }

\addplot[
    color=blue,
    mark=+,
    thick,
    dashed,
    mark size=3,
]
table {
-3.0 84.6033
-2.8 84.5521
-2.6 84.5032
-2.4 84.6431
-2.2 84.5052
-2.0 84.7461
-1.8 84.8835
-1.6 84.645
-1.4 84.7239
-1.2 84.6467
-1.0 84.6362
-0.8 84.6771
-0.6 84.6782
-0.4 84.6689
-0.2 84.6658
0.0 84.6581
0.2 84.6523
0.4 84.6662
0.6 84.6709
0.8 84.6673
1.0 84.6705
1.2 84.669
1.4 84.6787
};
\addlegendentry{\small $N=512$, $K=57$ - no UEID}

\end{axis}
\end{tikzpicture}
  \\
  \ref{ESlegend}
  \caption{Average percentage of estimated bits during the second decoding phase with early stopping when $L_{\max}=8$ and $C_2=5$.}
  \label{fig:avg_est}
\end{figure}

The impact of the devised early stopping criterion on the average number of estimated bits is shown in Fig.~\ref{fig:avg_est}, for $K=32$ and $K=57$. These results consider each of the $C_2$ candidates separately, since the number of candidates of length $N_1$ and $N_2$ in the second phase depends on the PMs received from the first phase, and thus on channel SNR. The solid curves have been obtained in cases the UE ID was sent through the considered code, while the dashed curves in cases it was not sent through the code.

\begin{itemize}
 \item For $N=256$ (curves with a circle marker), it is possible to observe the same behavior noted in \cite{Condo_COMML17} for $N=128$ as well. In case the UE ID was sent, as the channel conditions improve, the number of estimated bits increases until stabilizing at a maximum average value. This phenomenon can be explained by the fact that when the SNR is low, it is more likely that the codeword carrying the UE ID is not selected to be among the $C_2$ candidates. Thus the decoders in the second phase easily encounter ID bits different from the UE ID early in the decoding process. As the channel conditions improve, the codeword with the UE ID falls among the $C_2$ candidates with rising probability. Consequently, the decoder tasked with its decoding does not interrupt the process, reaching $100\%$ estimated bits, while the remaining $C_2-1$ decoders stop the decoding early, thus averaging the estimated bit percentage at a stable value ($67\%$ for $K=32$ and $61\%$ for $K=57$). The dashed curves show instead a stable value regardless of channel conditions: since among the $C_2$ candidates there is never one carrying the UE ID, all second phase decoders tend to stop the decoding early, at a percentage independent of the SNR, and mostly influenced by the position of bits assigned to the ID.
\item For $N=512$ (curves with a cross marker) a similar behavior to the $N=256$ case can be observed when the UE ID is not sent, with the average number of estimated bits stable at all the considered SNR values. On the other hand, when the UE ID is sent, the trend is different: at low SNR values, the percentage of estimated bits is very close to $100\%$. As the SNR value increases, the average starts to decrease, until it settles on a stable value.
This behavior is due to the fact that at low SNR, it is very unlikely that a codeword with $N=512$ is among the $C_2$ second phase candidates if the UE ID is not matching: the longer code length and lower rate contribute to a higher decoding reliability during the first phase, that allows to screen out unlikely candidates better than the $N=256$ case.
\end{itemize}

% 
% \begin{figure*}[t!]
%   \centering
%   \input{./AvgEstDistCRC256_27.tikz}
%   \\
%   \ref{ESlegend256_27}
%   \caption{Average percentage of estimated bits during the second decoding phase, with early stopping, $L_{\max}=8$, $C_2=5$.}
%   \label{fig:avg_est_RM3}
% \end{figure*}
% 
% 
% \begin{figure*}[t!]
%   \centering
%   \input{./AvgEstDistCRC256_49.tikz}
%   \\
%   \ref{ESlegend256_49}
%   \caption{Average percentage of estimated bits during the second decoding phase, with early stopping, $L_{\max}=8$, $C_2=5$.}
%   \label{fig:avg_est_RM3}
% \end{figure*}
% 
% 
% \begin{figure*}[t!]
%   \centering
%   \input{./AvgEstDistCRC512_27.tikz}
%   \\
%   \ref{ESlegend512_27}
%   \caption{Average percentage of estimated bits during the second decoding phase, with early stopping, $L_{\max}=8$, $C_2=5$.}
%   \label{fig:avg_est_RM3}
% \end{figure*}
% 
% 
% \begin{figure*}[t!]
%   \centering
%   \input{./AvgEstDistCRC512_49.tikz}
%   \\
%   \ref{ESlegend512_49}
%   \caption{Average percentage of estimated bits during the second decoding phase, with early stopping, $L_{\max}=8$, $C_2=5$.}
%   \label{fig:avg_est_RM3}
% \end{figure*}

% 
% \begin{figure*}[t!]
%   \centering
%   \input{./AvgEstRM3_FastSSCLSPC.tikz}
%   \caption{Average percentage of estimated bits with early stopping during the second phase, Fast-SSCL-SPC algorithm.}
%   \label{fig:avg_est_RM3_fast}
% \end{figure*}

\section{Hardware Architecture} \label{sec:HW}
\begin{figure}[t!]
  \centering
  \includegraphics[scale=0.55]{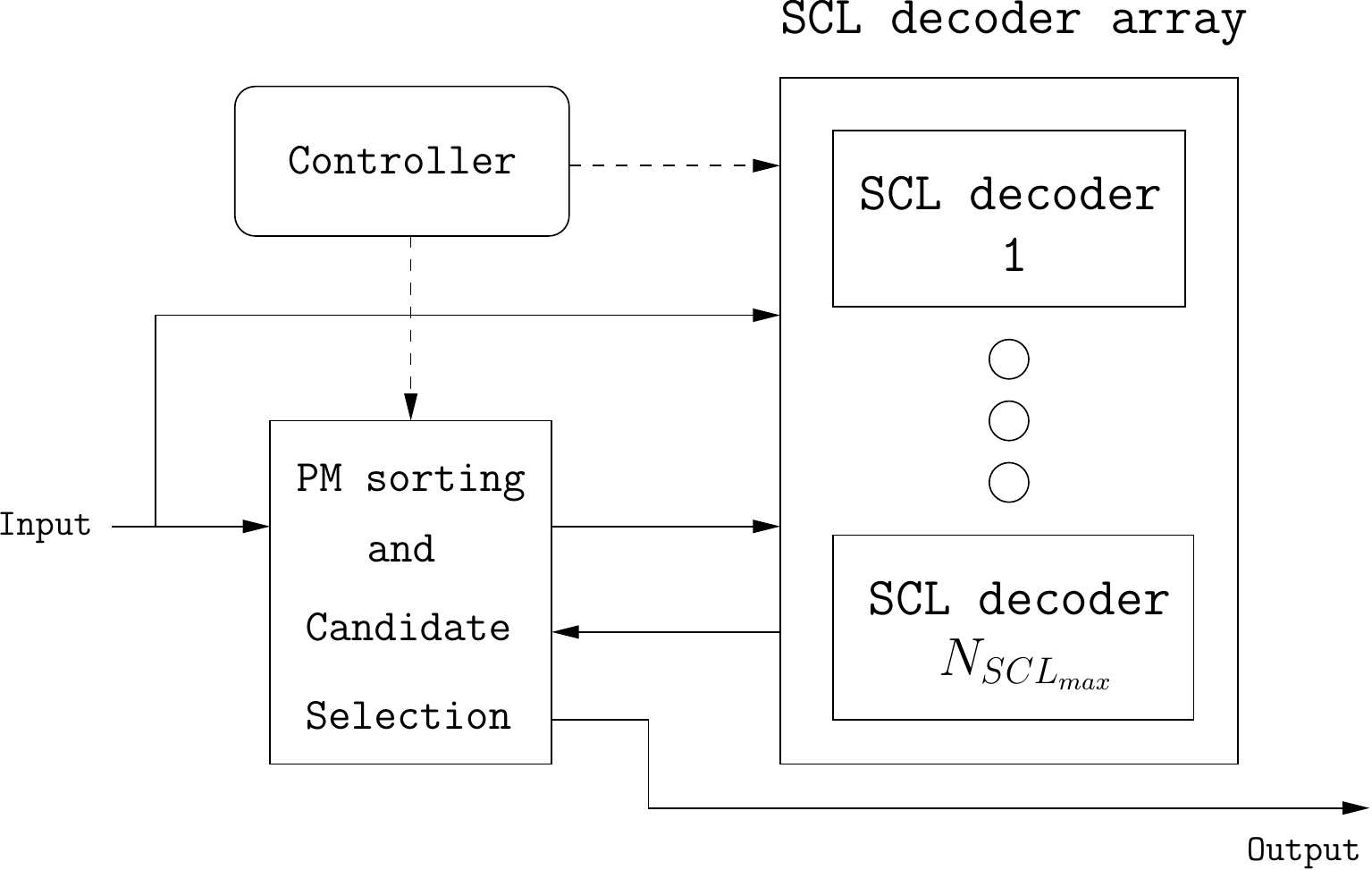}
  \caption{Polar codes blind detection system architecture.}
  \label{fig:BDarch}
\end{figure}

To evaluate the implementation cost of the devised blind detection scheme, we designed a decoder architecture that supports it, portrayed in Fig.~\ref{fig:BDarch}. An array of flexible list size SCL decoders handles both the first and second decoding phase. A dedicated module selects the $C_2$ candidates for the second phase according to the criteria described in Section~\ref{sec:blind}.

\subsection{Flexible list size SCL decoder}

We based our SCL decoder architecture on that of \cite{hashemi_SSCL_TCASI,hashemi_FSSCL}: the decoding process follows the one described in Section~\ref{sec:prel:PC} for a list size $L_{\max}$. Most of the datapath and memories are instantiated $L_{\max}$ times: multiple candidates are stored at the same time, with the best candidate being selected at the end of the decoding. While in \cite{hashemi_SSCL_TCASI,hashemi_FSSCL} the final candidate is selected according to a CRC check, in the proposed architecture no CRC is considered, and the validity of the final candidate is based on the matching ID and PM value.

The SC decoding tree is descended by computing (\ref{eq1}) and (\ref{eq2}) at each stage $s$, with priority being given to left branches. These calculations are performed by $L_{\max}$ parallel sets of $P$ processing elements (PEs), with $P$ being a power of $2$. In the stages for which $2^s>2P$, the operations in (\ref{eq1}) and (\ref{eq2}) are performed over $2^s/(2P)$ steps, while a single step is needed otherwise. Internal memories store the updated LLR values between stages.

PEs get two LLR values as input, and concurrently compute both $\alpha^{\text{l}}$ and $\alpha^{\text{r}}$ according to (\ref{eq1}) and (\ref{eq2}), respectively. The correct output is selected depending on the index of the leaf node to be estimated. When a leaf node is reached, the decoder controller module identifies the leaf node as either an information bit or a frozen bit. If a frozen bit is found, the paths are not split, and the bit is estimated only as $0$, and the $L$ memories are updated with the same bit or LLR values. Instead, in case of an information bit, both $0$ and $1$ are considered, so that paths are split, and the PMs updated for the $2L$ candidates according to (\ref{eq7}). Afterwards, the PMs are sorted, identifying the $L$ surviving paths.

All memories in the decoder are registers, enabling the internal LLR and $\beta$ values to be read, updated by the PEs, and written back in a single clock cycle. At the same time, the paths are either updated or split and updated, and the new PMs computed. In the following clock cycle, in case the paths were split, the PMs are sorted and the surviving paths selected.

Codes with different code lengths can be decoded by storing the appropriate memory offsets for every considered code in a dedicated memory.

This baseline decoder has been modified to better fit the needs of the proposed blind detection scheme. In order to maximize resource sharing, the SCL decoder has been sized for $L_{\max}>L_1$, and the effective list size can be selected through a dedicated input.
The $L_{\max}-L_1$ paths that are not used in the first decoding phase are used to decode up to $\left\lfloor(L_{\max}-L_1)/L_1\right\rfloor$ additional candidates at the same time. In order to exploit the unused paths, additional functional modules are necessary.
\begin{itemize}
 \item The baseline decoder uses a single memory to store the channel LLR values, sharing it among the different paths. If different codewords have to be decoded at the same time, the channel memory needs to be instantiated not once, but $\left\lfloor L_{\max}/L_1\right\rfloor$ times.
 \item The decoder relies on sorting and selection logic that identifies the surviving $L_{\max}$ ones after paths are split. To support the parallel decoding of $\left\lfloor L_{\max}/L_1\right\rfloor$ candidates, as many sorting and selection modules targeting the selection of $L_1$ paths out of $2L_1$ are instantiated.
\end{itemize}
If $L_1=1$ is selected, the path splitting and PM sorting steps are bypassed, reverting decoders to the standard SC case. Since a single set of SCL decoders can handle both decoding phases, the total number of decoders is $N_{\text{SCL}_{\max}}$ (see Fig. \ref{fig:BDarch}). However, the effective number of decoders for the first decoding phase is $N_{\text{SCL}_1}=N_{\text{SCL}_{\max}}\times \left\lfloor L_{\max}/L_1\right\rfloor$.

The early stopping technique described in Section~\ref{sec:blind} has been also implemented. The decoder receives as input the position of the ID bits and the value of the UE ID: every time a bit in an ID position is estimated, the bit value is compared to the expected UE ID bit. All paths whose estimated bit does not match the UE ID bit are deactivated. This operation is performed after the $L$ surviving paths have been selected, in order not to force the survival of unlikely paths and increase the FAR. In case all paths have been deactivated, the decoding is stopped. The early stopping logic can be activated and deactivated by means of a dedicated control signal. Since the same hardware is used for both decoding phases, early stopping is enabled only during the second one.

\subsection{PM sorting and candidate selection}

\begin{figure*}[t!]
  \centering
  \includegraphics{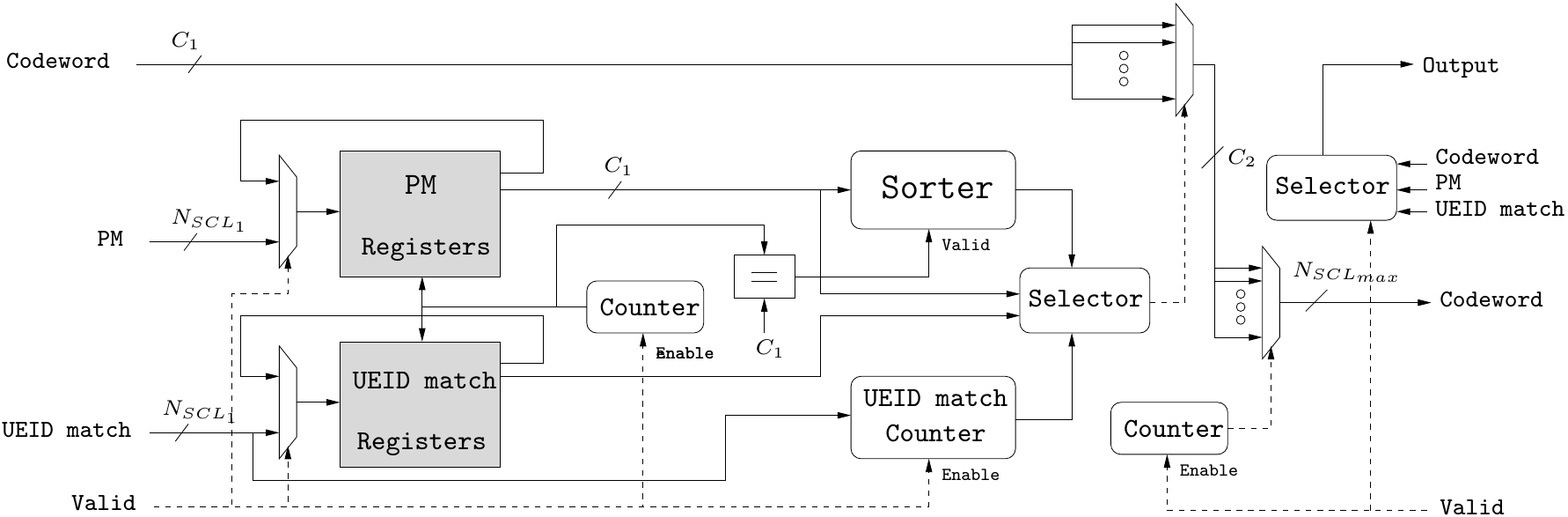}
  \caption{PM sorting and candidate selection architecture.}
  \label{fig:C2sel}
\end{figure*}

Fig.~\ref{fig:C2sel} depicts the architecture of the PM sorting and candidate selection block. It processes the output of the first decoding phase to select the $C_2$ candidates for the second phase, and selects the overall system output based on the results from the second phase. For each of the $N_{\text{SCL}_1}$ first phase decoders, a PM and a flag signalling a UE ID match are received. They are stored every time the respective \texttt{Valid} signal is risen by the decoder. The \texttt{Valid} signal is also used as an enable for the PM and UE ID match register address counter, and for the counter keeping track of how many codewords had a matching UE ID after the first phase. When all the $C_1$ candidates have gone through the first decoding phase, a \texttt{Valid} signal is issued to the sorter module, that receives as input all the stored PMs. The sorter module returns the $C_2$ minimum PMs in as many clock cycles: each PM is compared to all the others, and a single clock cycle is necessary to identify the minimum one, that is excluded from the subsequent comparison. When the $C_2$ minima have been found, the selector module considers how many candidates had a matching UE ID after the first phase, and selects the $C_2$ candidates for the second phase among them and those with the minimum PM values. The $C_2$ candidates are sent to the $N_{\text{SCL}_{\max}}$ decoders by means of a dedicated counter. Returning PMs and UE ID match flags are received and compared by another selector: when all $C_2$ candidates have been decoded, the selected codeword, if any, is output.

\section{Implementation Results} \label{sec:impl}

The architecture proposed in Section~\ref{sec:HW} has been described in VHDL and synthesized in TSMC 65~nm CMOS technology. Table~\ref{tab:asic} reports the synthesis results for the architecture sized for a maximum code length $N_{\max}=512$, a maximum list size $L_{\max}=8$, $C_2=5$, and a target frequency $f=1$~GHz. Various $N_{\text{SCL}_{\max}}$ values have been considered, leading to different latencies and area occupations. Since during the first decoding phase $L_1=2$, the effective number of decoders $N_{\text{SCL}_1}$ is equal to $4N_{\text{SCL}_{\max}}$, even if only $N_{\text{SCL}_{\max}}$ are physically instantiated.
Regarding the area, the $N_{\text{SCL}_{\max}}$ SCL decoders contribute to the majority of the complexity, ranging from $97.8\%$ when $N_{\text{SCL}_{\max}}=1$ to $99.7\%$ when $N_{\text{SCL}_{\max}}=5$. The logic complexity of the PM sorting and candidate selection module remains almost unchanged at the variation of $N_{\text{SCL}_{\max}}$, being mainly affected by $C_1$ and $C_2$. Memories have been synthesized with registers only, without the use of RAM, and account for $36\%$ of the total area occupation.

The worst case latency of the proposed blind detection system can be found as 
\begin{align} \label{eq:lat2}
\begin{split}
 T_{\text{bd}}=&\left\lceil \frac{C_1}{N_{\text{SCL}_1}} \right\rceil \left(\frac{T^1_{\text{SCL}}}{2}+\frac{T^2_{\text{SCL}}}{2}\right)\\
 &+ T_{\text{sort}} + \left\lceil \frac{C_2}{N_{\text{SCL}_{\max}}} \right\rceil \max\left(T^1_{\text{SCL}},T^2_{\text{SCL}}\right)~ \text{,}\\
\end{split} 
\end{align}
where $T^1_{\text{SCL}}$ and $T^2_{\text{SCL}}$ are the SCL decoding latencies for codes of length $N_1$ and $N_2$, respectively, while $T_{\text{sort}}$ is the number of time steps required to sort the PM of the first decoding phase and obtain the $C_2$ candidates out of the $C_1$ candidate locations. Also, it is worth remembering that for the proposed architecture, $N_{\text{SCL}_1}=\left\lfloor L_{\max}/L_1\right\rfloor \times N_{\text{SCL}_{\max}}$.
The SCL decoding latency can be found as \cite{balatsoukas}
\begin{equation*}
T^x_{\text{SCL}}=2N_x+K_x+16-2 \text{,}
\end{equation*}
for $x\in\{1,2\}$.
From the results presented in Table \ref{tab:asic}, it is possible to see that even when considering the relatively old 65~nm technology node, the $16\mu$s worst case latency target can be reached with a single SCL decoder running at a frequency of $1$~GHz, while $N_{\text{SCL}_{\max}}=5$ guarantees a worst case latency of $3.6\mu$s, meeting the $4\mu$s target as well.

However, considering only the worst case latency is indeed an unrealistic scenario. To begin with, while there is no guarantee on how the $C_2$ candidates are distributed among $N_1$ and $N_2$, simulation results have shown that we can expect the $C_2$ candidates either to favor the shorter code length, or to be equally divided between $N_1$ and $N_2$ candidates. Thus, the factor $$\left\lceil\frac{C_2}{N_{\text{SCL}_{\max}}} \right\rceil\max\left(T^1_{\text{SCL}},T^2_{\text{SCL}}\right)$$ in (\ref{eq:lat2}), that represents the contribution of the second decoding phase, could be better expressed as: 
\begin{equation*}
\left\lceil\frac{\left\lceil C_2/2\right\rceil}{N_{\text{SCL}_{\max}}} \right\rceil T^1_{\text{SCL}} + \left\lceil\frac{C_2-\left\lceil C_2/2\right\rceil}{N_{\text{SCL}_{\max}}} \right\rceil T^2_{\text{SCL}}~.
\end{equation*}
Note that this is still a conservative assumption, since it entails the $C_2$ candidates equally divided among the two code lengths. We can refine this assumption by taking in account the effect of early stopping. We can approximate the latency reduction with a multiplicative factor $E^x$ associated to  $T^x_{\text{SCL}}$. Consequently, the average latency of the blind detection system, for $N_{\text{SCL}_{\max}}<C_2$, can be computed as
\begin{align} \label{eq:lat3}
\begin{split}
 T_{\text{bd}}&=\left\lceil \frac{C_1}{N_{\text{SCL}_1}} \right\rceil \left(\frac{T^1_{\text{SCL}}}{2}+\frac{T^2_{\text{SCL}}}{2}\right) + T_{\text{sort}}\\
 & + \left\lceil\frac{\left\lceil C_2/2\right\rceil}{N_{\text{SCL}_{\max}}} \right\rceil T^1_{\text{SCL}}E^1 + \left\lceil\frac{C_2-\left\lceil C_2/2\right\rceil}{N_{\text{SCL}_{\max}}} \right\rceil T^2_{\text{SCL}}E^2~ \text{,}\\
\end{split} 
\end{align}
while for $N_{\text{SCL}_{\max}}\ge C_2$ it becomes
 \begin{align} \label{eq:lat4}
 \begin{split}
T_{\text{bd}}=\left\lceil \frac{C_1}{N_{\text{SCL}_1}} \right\rceil  & \left(\frac{T^1_{\text{SCL}}}{2}+\frac{T^2_{\text{SCL}}}{2}\right) + T_{\text{sort}}\\
  & + \max\left(T^1_{\text{SCL}}E^1,T^2_{\text{SCL}}E^2\right)~ \text{.}\\
 \end{split}  
 \end{align}

Considering the number of UEs connected to the shared channel, blind detection is dominated by instances in which a particular UE ID is not sent. Thus, we can set $E^x$ as the fraction of bits expressed by the dashed curves in Fig.~\ref{fig:avg_est}. The average latency results in Table~\ref{tab:asic} show substantial reduction with respect to the worst case latency case, within a more realistic framework. Even within the 65~nm technology node, with $N_{\text{SCL}_{\max}}\ge 4$, the average latency is below $4\mu$s. With the latest technology nodes, a substantially higher frequency will be easy to achieve, along with proportionally smaller area occupation. It is consequently safe to assume that the $4\mu$s worst case latency target can be easily met for $N_{\text{SCL}_{\max}}\ge3$, and the average latency with $N_{\text{SCL}_{\max}}\ge2$.

\begin{table}[tbp!]
  \centering
  %\scriptsize
    \caption{TSMC CMOS 65~nm blind detection scheme synthesis results for $L_{\max}=8$, $P=64$, $C_2=5$, and $f=1$~GHz.}
    		\setlength{\extrarowheight}{1.8pt}
  \begin{tabular}{cccccc}
\toprule
    \multirow{2}{*}{$N_{\text{SCL}_{\max}}$} & Area & \multicolumn{2}{c}{Worst case latency} & \multicolumn{2}{c}{Average latency}\\
    & [mm$^2$] & [clock cycles] & [$\mu$s] & [clock cycles] & [$\mu$s]\\
\cmidrule(lr){1-1}
\cmidrule(lr){2-2}
\cmidrule(lr){3-4}    
\cmidrule(lr){5-6}
    %1.326
    $1$ & $1.555$ & $14720$ & $14.7$ & $11843$ & $11.8$\\ %part1=11*(840)+5  part2=3*583*0.426 + 2*1095*0.846  3*249+ 2*927
    $2$ & $3.086$ & $8330$ & $8.3$ & $6470$ & $6.5$\\ %part1=6*(840)+5  part2=2*583*0.426 + 1*1095*0.846
    $3$ & $4.596$ & $5555$ & $5.6$ & $4541$ & $4.5$\\ %part1=4*(840)+5  part2=1*583*0.426 + 1*1095*0.846
    $4$ & $6.117$ & $4710$ & $4.7$ & $3696$ & $3.7$\\ %part1=3*(840)+5  part2=1*583*0.426 + 1*1095*0.846
    $5$ & $7.654$ & $3620$ & $3.6$ & $3451$ & $3.5$\\ %part1=3*(840)+5  part2= + 1*1095*0.846
\bottomrule
	\end{tabular}
  \label{tab:asic}
\end{table}

%    Type    Instances     Area    Area %
% ----------------------------------------
% sequential     49573  433166.400   28.5
% inverter       42646   96331.320    6.3
% buffer          5628   17235.360    1.1
% logic         318672  973766.160   64.0
% ----------------------------------------
% total         416519 1520499.240  100.0
%+34269

\section{Conclusion} \label{sec:conc}
In this work, we propose a polar codes blind detection scheme. The candidates go through a first, coarser decoding phase, that helps to select a few of them for a second, finer decoding phase. An early stopping criterion is proposed for the second phase, to reduce average latency. We evaluate the effectiveness of the blind detection scheme, and propose an architecture to implement it. It is based on an SCL decoder with tunable list size, that can be used for both decoding stages. The architecture is synthesized and implementation results are reported for various system parameters. The reported area occupation and latency, obtained in 65~nm CMOS technology, are able to meet 5G requirements, and are guaranteed to meet them with even less resource usage in the latest technology nodes.

\bibliographystyle{IEEEtran}
%\bibliography{IEEEabrv,refs}

\end{document}